\documentclass[aps,prb,preprint]{revtex4}
\usepackage{epsfig}
\begin{document}
\author{Zaher~Salman}
\address{Technion-Israel Institute of Technology, Physics\\
Department, Haifa 32000, Israel }
\title{Theoretical $T_1$ Calculation for Isotropic High Spin Molecules}
\begin{abstract}
We calculate the molecular-spin ($S$), temperature ($T$), and field
($H$) dependence of $1/T_1$ for a local magnetic probe coupled to an
isotropic high spin molecule, based on spin-phonon interaction. We
compare the calculation to recent NMR and $\mu$SR experiments in
CrCu$_6$ ($S = 9/2$), CrNi$_6$ ($S = 15/2$) and CrMn$_6$ ($S =
27/2$). Although we can account for the high and intermediate
temperature regimes, the calculation is fundamentally different from
the data at $T \longrightarrow 0$. Since $1/T_1$ must be due to
coupling of the molecular spin to an external heat bath, and since
phonon contribution is ruled out at low $T$, we conclude that at these
temperatures hyperfine interactions must play an important role in the
molecular spin dynamic.
\end{abstract}

\maketitle

\section{Introduction}
High spin molecules (HSM) consist of clusters of metal ions, they are
ordered in a crystal lattice, and coupled by Heisenberg ferromagnetic
or anti-ferromagnetic interactions with coupling constant $J$, only
between spins $\vec{S}_{i}$ in the molecule. In this paper we study a
family of isotropic high spin molecules. We calculate the spin lattice
relaxation $1/T_1$ in these molecules, and compare it to experimental
measurements\cite{Salman02PRB} to determine the origin of the observed
spin dynamics of the molecules. We find that at high temperature the
molecular spin dynamics is driven by thermal activation (spin-phonon
interaction). However at low temperatures the molecular spin dynamics
is dominated by hyperfine interactions between the molecular and
nuclear spin. 

These calculations can be easily applied to different kinds of high
spin molecules, such as
Mn$_{12}$\cite{Lis80ACS,Thomas96N,Freidman96PRL} and
Fe$_8$\cite{Sangregorio97PRL}, where quantum tunneling of the
magnetization (QTM) is observed
\cite{Thomas96N,Freidman96PRL,Sangregorio97PRL}. Comparison between
experimental measurements\cite{Ueda02PRB,Lascialfari98PRL} and the calculation
can help determine which interactions induce the spin tunneling in
these molecules.

\begin{figure}[h]
\centerline{\epsfysize=5.0cm \epsfbox{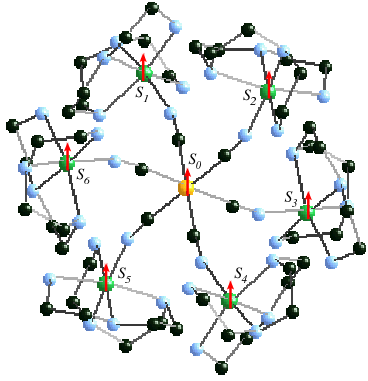}}
\caption{The magnetic core of the CrNi$_6$ molecule, $S_0$ is the spin of the Cr ion and $S_1,S_2,\cdots,S_6$ are the spins of the Ni ions.} \label{CrNi6core}
\end{figure}

In the isotropic molecules a Cr(III) ion is surrounded by six cyanide
ions, each bonded to a Cu(II), Ni(II) or Mn(II) ion. The coordination
sphere of Cr and Cu/Ni/Mn can be described as a slightly distorted
octahedral. For convenience we will refer to these molecules as
CrCu$_6$, CrNi$_6$ and CrMn$_6$, respectively. The Hamiltonian of the
isotropic high spin molecules at zero field can be written as
\begin{equation} \label{H0}
{\cal H}_0=-J \sum_{i=1}^6 \vec{S}_0 \cdot \vec{S}_i
\end{equation}
where $\vec{S}_0$ is the spin of the central Cr ion (with $S=3/2$),
$i$ runs over the peripheral Cu, Ni or Mn ions (with $S=1/2$, $S=1$ or
$S=5/2$), which are coupled to the central Cr ion, with coupling
constant $J$ (e.g. see Figure~\ref{CrNi6core}). At temperatures lower
than $J$ only the ground spin state $S=9/2$, $S=15/2$ or $S=27/2$ is
populated.

The Hamiltonian ${\cal H}_0$ is isotropic therefore the total spin $S$
and the spin in the $z$ direction $m$ are good quantum numbers, and
the eigenfunctions of ${\cal H}_0$ can be written as $|S,m>$. However,
at very low temperatures $T \ll J$, the degeneracy of the ground state
is removed by additional anisotropic perturbation on ${\cal H}_0$. 
Such perturbation that does not commute with ${\cal H}_0$ can cause
transitions between the different spin states $|S,m>$
\cite{Politi95PRL,Barra97PRB,Barra98JMMM,Fort98PRL,Caneschi98JMMM,Villain94EL,Prokofev98PRL,Salman00,Salman02PRB,Garanin97PRB,Garg98PRL},
and induce the observed spin dynamics \cite{Salman00,Salman02PRB}.

The anisotropic term in the Hamiltonian can be written as the sum
${\cal H}`={\cal H}_c+{\cal H}_n$, where ${\cal H}_c$ commutes with
the Hamiltonian ${\cal H}_0$, while ${\cal H}_n$ does not. These terms
may be a result of dipolar interaction between neighboring spins
\cite{Politi95PRL,Garanin97PRB}, spin-phonon interaction
\cite{Villain94EL,Garg98PRL}, nuclear fluctuations
\cite{Prokofev98PRL}, high order crystal field terms
\cite{Sangregorio97PRL,Politi95PRL,Barra97PRB,Barra98JMMM,Fort98PRL,Caneschi98JMMM}
or small anisotropy in the coupling $J$ between spins
\cite{Miyashita01condmat}.

In order to calculate the value of the spin lattice relaxation time
$T_1$ in the isotropic HSM we diagonalize the Hamiltonian ${\cal
H}_0+{\cal H}_c$ in Section~\ref{exact}. In Section~\ref{susfit} we
show that we can calculate the magnetization and susceptibility of the
compounds using the eigenvalues and eigenfunctions that we obtain. In
Section~\ref{T1calculation} we use the eigenvalues and eigenfunctions
of the Hamiltonian to calculate $T_1$, taking ${\cal H}_n$ as a
perturbation. The perturbation introduces a finite lifetime for the
different spin states $|S,m>$ and induces spin dynamics, or
transitions between the different spin states.

\section{Exact diagonalization of the Hamiltonian} \label{exact}
To calculate the eigenvalues of the Hamiltonian (\ref{H0}) we write it
 in the form
\begin{equation}
{\cal H}_0=-J \vec{S}_0 \cdot \sum_{i=1}^6 \vec{S}_i.
\end{equation}
Using $\vec{S}=\vec{S}_0+\sum_{i=1}^6 \vec{S}_i$ and $\vec{S}_t=\sum_{i=1}^6 \vec{S}_i$ one can write
\begin{equation}
{\cal H}_0 = -J\vec{S}_0 \cdot \vec{S}_t = -\frac{J}{2} \left( \vec{S}^2 -\vec{S}_t^2 - \vec{S}_0^2 \right).
\end{equation}
The energy eigenvalues of the sates $|S,m> \equiv |S_0,S_t,S,m>$ are
\begin{equation}
E_{\bf{S}} \equiv E(S,S_t,S_0)=-\frac{J}{2} \left( S(S+1)-S_t(S_t+1)-S_0(S_0+1) \right)
\end{equation}
where we use the notation ${\bf S}$ for the set of numbers
$(S,S_t,S_0)$. The degeneracy of the state $|S,m>$ is the degeneracy
of the value of $S_t$. When an external magnetic field $H$ is applied
the Zeeman term should be added to ${\cal H}_0$, and the full
Hamiltonian becomes
\begin{equation} \label{Hfull}
{\cal H}= {\cal H}_0 - g \mu_B H S_z,
\end{equation}
for which the eigenfunctions $|S,m>$ are not changed and the
eigenvalues are
\begin{equation}
E_{{\bf S},m}=-\frac{J}{2} \left( S(S+1)-S_t(S_t+1)-S_0(S_0+1) \right) -g \mu_B H m.
\end{equation}

\begin{figure}[h]
\centerline{\epsfysize=6.0cm \epsfbox{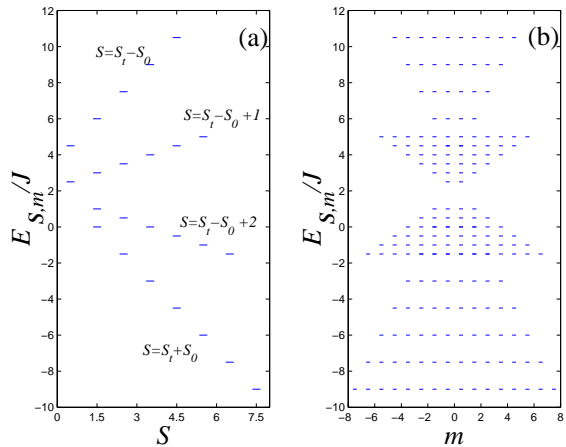}}
\caption{The energy levels of the Hamiltonian ${\cal H}_0$ at zero field for CrNi$_6$ as a function of (a) the total spin $S$ and (b) the spin in the $z$ direction $m$.} \label{energy}
\end{figure}

In Figure~\ref{energy} we present the energy levels of CrNi$_6$ at
zero magnetic field $H=0$ as a function of the total spin $S$ (a) and
as a function of the spin in $z$ direction $m$ (b).

\section{Susceptibility Calculation} \label{susfit}
\begin{figure}[h]
\centerline{\epsfysize=4.0cm \epsfbox{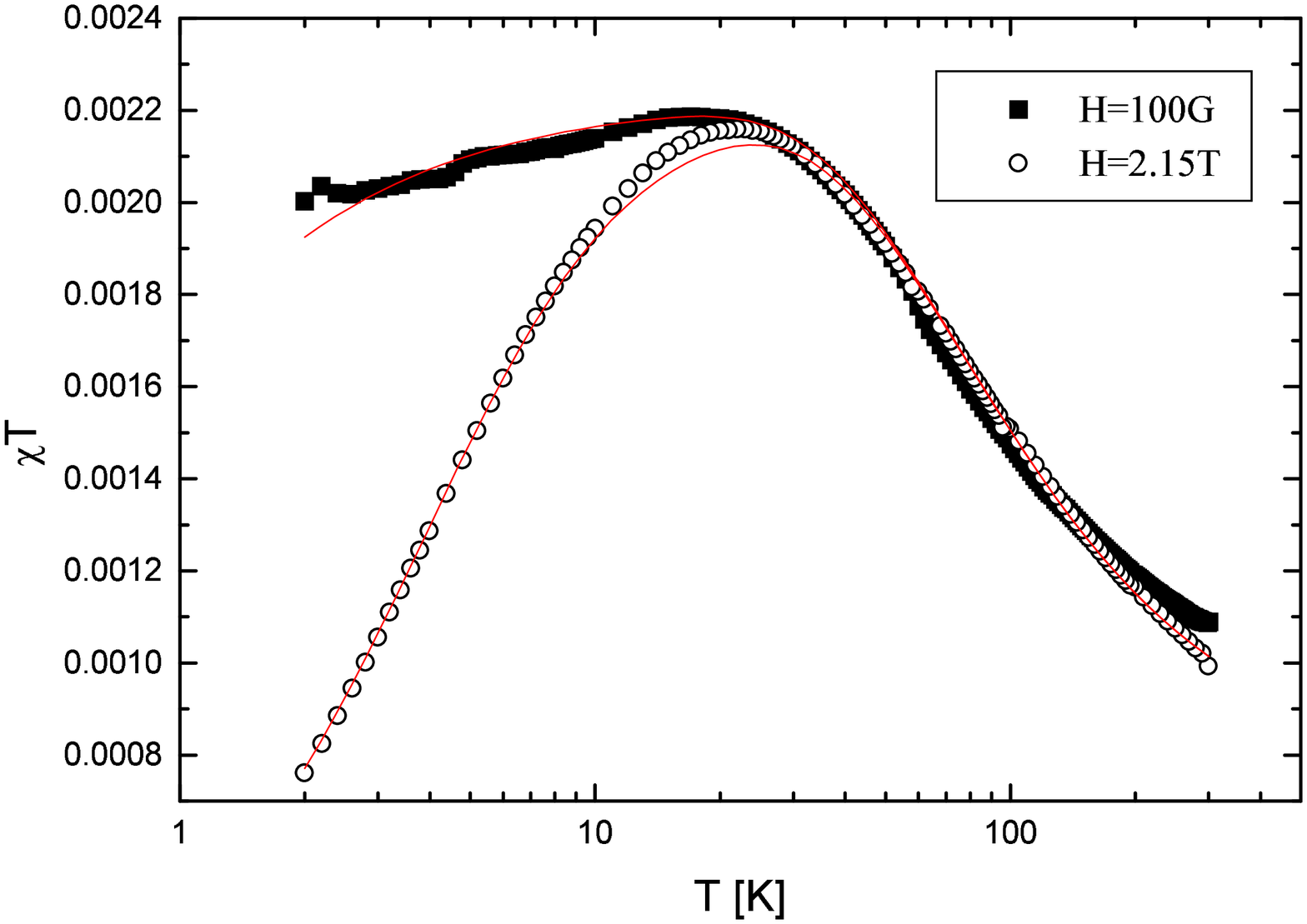} \hspace{-1.0cm} (a)
\epsfysize=4.0cm \epsfbox{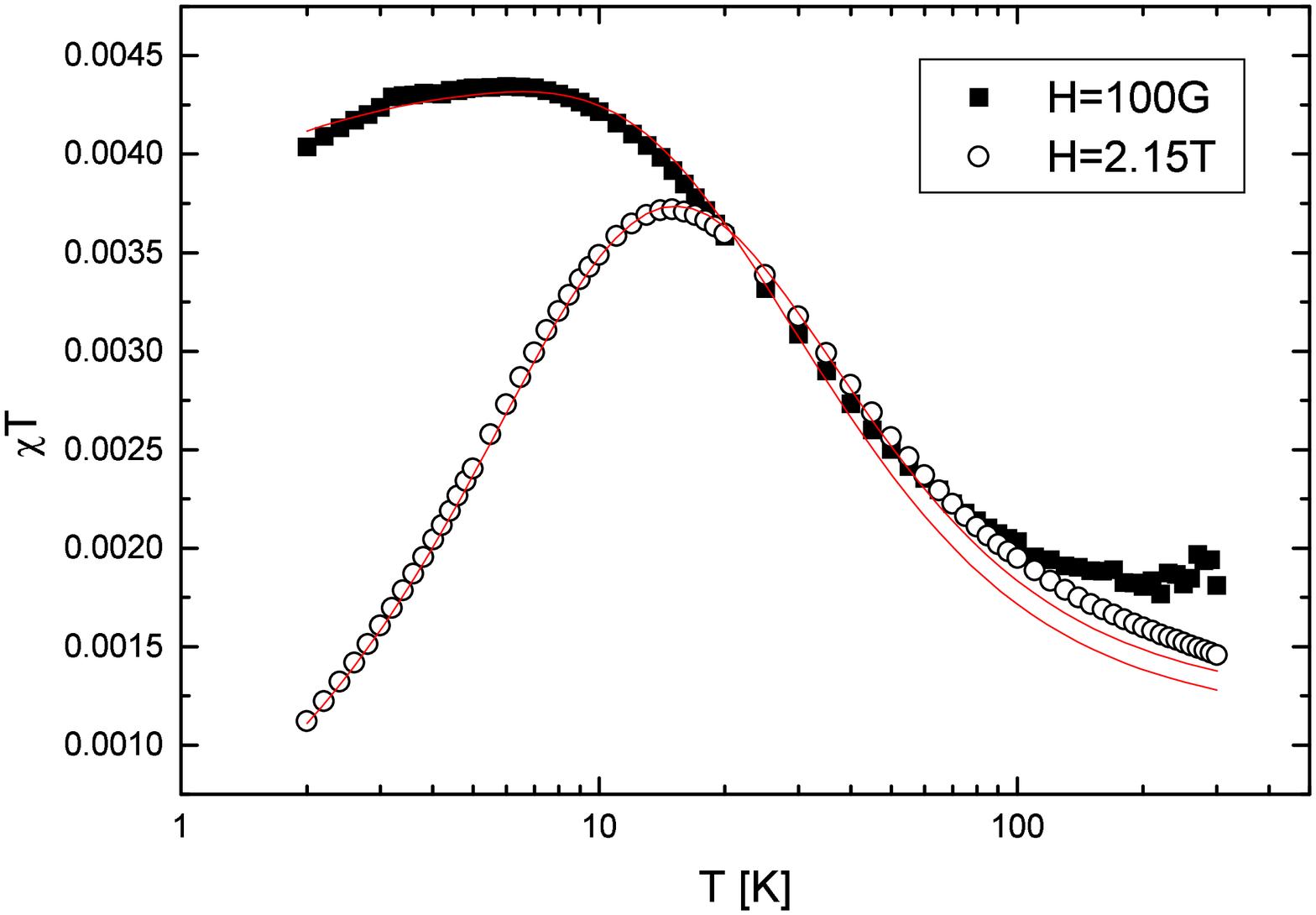} \hspace{-1.0cm} (b)
\epsfysize=4.0cm \epsfbox{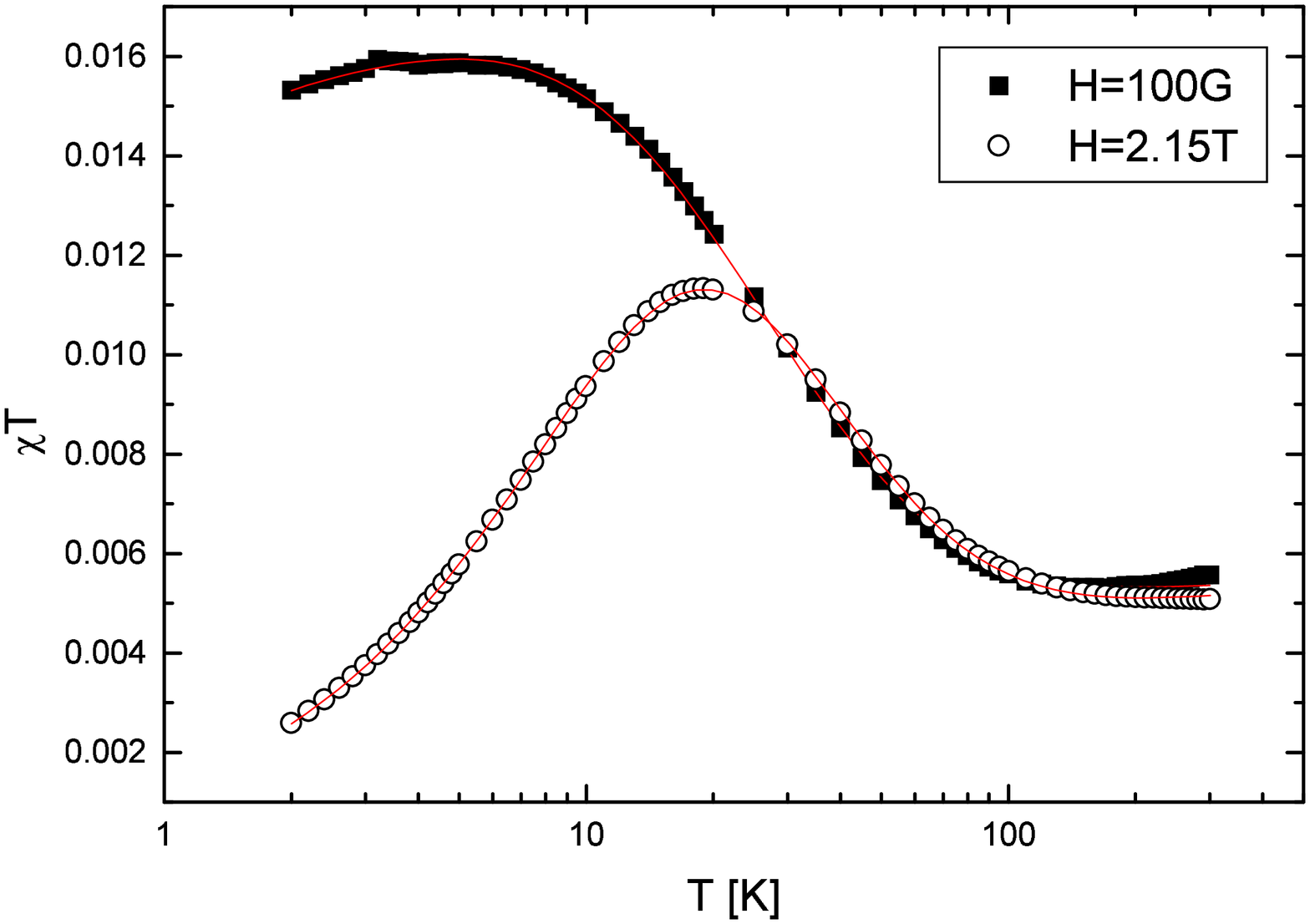}\hspace{-1.0cm}(c)}
\caption{The susceptibility multiplied by temperature as a function of temperature at two different external fields, measured in (a) CrCu$_6$ (b) CrNi$_6$ and (c) CrMn$_6$. The solid lines are fits to the theoretical expectation (see text).} \label{chit}
\end{figure}

The susceptibility can be calculated as a function of
temperature and external field using the eigenstates and eigenvalues
that we calculated in the previous section. First we calculate the
magnetization
\begin{equation}
M(T)=<S_z>=\sum_{|S,m>} \frac{m e^{-\frac{E_{{\bf S},m}}{T}}}{{\cal Z}}
\end{equation}
where ${\cal Z}=\sum_{|S,m>} \exp \left( - \frac{E_{{\bf S},m}}{T}
\right)$ is the partition function. The measured
susceptibility\cite{Salman02PRB} is defined as 
\begin{equation} \label{chicalc}
\chi(T)=\frac{M(T)}{H}.
\end{equation}
In Figure~\ref{chit} we fit the experimental measurement of $\chi T$
as a function of temperature in (a) CrCu$_6$, (b) CrNi$_6$ and (c)
CrMn$_6$, at fields $H=100$~G and $2.15$~T, to the calculated value of
$\chi T$ from the Hamiltonian (\ref{Hfull}) and
Eq.~(\ref{chicalc}). From the fit one obtains the coupling constants
$J_{\rm Cr-Cu}=77$~K, $J_{\rm Cr-Ni}=24$~K and $J_{\rm Cr-Mn}=-11$~K
and no anisotropy is observed in all three molecules, within the fitting
accuracy.
This indicates that the high spin molecules can be well described by
the Hamiltonian (\ref{Hfull}), and that the molecules are indeed
isotropic.

\section{$T_1$ Calculation} \label{T1calculation}
The spin lattice relaxation time $T_1$ was measured
\cite{Salman02PRB} in CrCu$_6$, CrNi$_6$ and CrMn$_6$ using $\mu$SR
(where the local probes are polarized muons) and proton-NMR. Assuming
for simplicity an isotropic interaction between the probe $\vec{I}$
(muon or proton) and the local electronic spins $\vec{S}$ of the whole molecule
\begin{equation}
{\cal H}_{IS} = A \vec{I} \cdot \vec{S}
\end{equation}
the spin lattice relaxation is given by \cite{White}
\begin{equation} \label{T1Eq}
\frac{1}{T_1}=\frac{A^2}{2} \int_{- \infty}^{\infty} \left<
S_{-}(t)S_{+}(0) \right> e^{i \omega t} dt
\end{equation}
where $\omega=\gamma H$ is the probe's Larmor frequency in an
external field $H$. The value of
\[
\left< S_-(t) S_+(0) \right>=\left< e^{-i {\cal H}t/\hbar} S_- e^{i
{\cal H}t/\hbar} S_+ \right>
\]
for the eigenstates $|S,m>$ of the Hamiltonian ${\cal H}$ in
Eq. (\ref{Hfull}) is,
\[
\left< S_-(t) S_+(0) \right>= \sum_{|S,m>} \left( S(S+1)-m(m+1) \right)
e^{i \left( \frac{E_{{\bf S},m+1}-E_{{\bf S},m}}{\hbar} \right) t}
\frac{e^{-\frac{E_{{\bf S},m}}{T}}}{\cal Z}
\]
where ${\cal Z}$ is the partition function.
This result is correct for a system with levels of infinitely long
lifetime. However, assuming a Lorentzian broadening of the levels, due
to the non-commuting additional terms in the Hamiltonian, ${\cal H}_n$,
we arrive at
\begin{equation}
\left< S_-(t) S_+(0) \right>= \sum_{|S,m>}
\frac{e^{-\frac{E_{{\bf S},m}}{T}}}{\cal Z} \left( S(S+1)-m(m+1) \right)
e^{-\frac{t}{\tau_{S,m}}} e^{i \left( \frac{E_{{\bf S},m+1}-E_{{\bf S},m}}{\hbar}
\right)t}
\end{equation}
where $\tau_{S,m}$ is the lifetime (or the inverse of the broadening)
of the level $|S,m>$. Hence according to Eq. (\ref{T1Eq})
\begin{equation}
\frac{1}{T_1}=\frac{A^2}{2} \sum_{|S,m>}
\frac{e^{-\frac{E_{{\bf S},m}}{T}}}{\cal Z} \left( S(S+1)-m(m+1) \right)
\int_{- \infty}^{\infty} e^{- \frac{t}{\tau_{S,m}}} e^{i \omega' t} dt
\end{equation}
where $\omega'=\omega+(E_{{\bf S},m+1}-E_{{\bf S},m})/\hbar$. The
energy difference in our case of isotropic molecules is
$E_{{\bf S},m+1}-E_{{\bf S},m} = -g \mu_B H$, therefore
$\omega'=\omega-\frac{g \mu_B}{\hbar}H$ and
\begin{equation} \label{T1calc}
\frac{1}{T_1}= \frac{A^2}{2{\cal Z}} \sum_{|S,m>} \left(S(S+1)-m(m+1) \right) \left( \frac{ \tau_{S,m} e^{-\frac{E_{{\bf S},m}}{T}}}{1 +\omega'^2 \tau_{S,m}^2} \right),
\end{equation}
in contrast to what was obtained in \cite{Lascialfari98PRL}, where the
term $\left(S(S+1)-m(m+1) \right)$ is missing in Eq.~(\ref{T1calc})
and $\omega'$ is replaced by $\omega=\gamma H$.

\begin{figure}
\centerline{\epsfysize=6.0cm \epsfbox{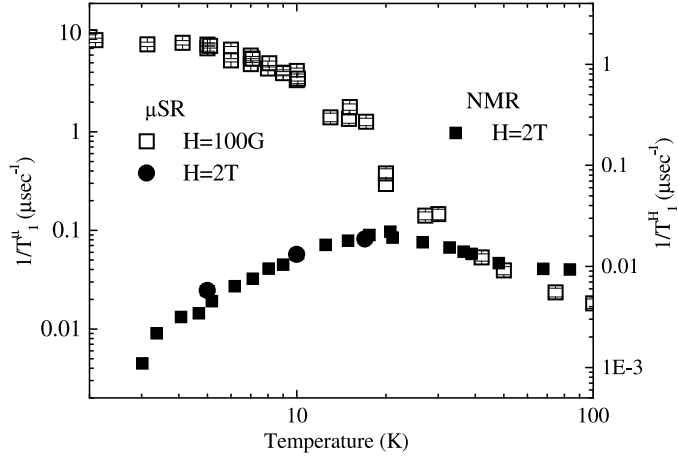}}
\caption{The values of the spin lattice relaxation measured by $\mu$SR ($1/T_1^\mu$), compared to those measured by proton NMR ($1/T_1^H$) after scalling, at the same external field $H=2.15$.}\label{NMRmuSR}
\end{figure}

This result is important for understanding of the comparison between
the spin lattice relaxation rate measured by $\mu$SR and NMR at
$H=2.15$~Tesla, shown in Figure~\ref{NMRmuSR}.
 The field dependence of $1/T_1$ comes from $\omega'=(\gamma-g
\mu_B/\hbar) H$. Since the gyromagnetic ratio $\gamma$ of the probe
(muon or nucleus) is much smaller than $g \mu_B / \hbar$ (electronic
gyromagnetic ratio), one can write $\omega' \simeq -g \mu_B H/\hbar$
which does not depend on the value of $\gamma$. Therefore the field
dependence of the spin lattice relaxation rate $1/T_1$ is independent
of the value of $\gamma$, and depends only on the value of $A$. This
explains the fact that the spin lattice relaxation rate measured by
proton-NMR can be scaled to match that measured by $\mu$SR {\em at the
same external field}, and not at the same Larmor frequency of the
probe (see Figure~\ref{NMRmuSR}). However, Eq. (\ref{T1calc}) is valid
assuming that the perturbation ${\cal H}_n$ is smaller than the Zeeman
splitting, $E_{{\bf S},m+1}-E_{{\bf S},m}$, i.e. at high fields where
$g \mu_B H \gg {\cal H}_n$.  When the Zeeman splitting is smaller than
${\cal H}_n$, $\omega'$ should be replaced by $\omega=\gamma H$, in
Eq.  (\ref{T1calc}).

The lifetime $\tau_{S,m}$ of the level $|S,m>$ can be expressed in
terms of transition probability from the state $|S,m>$ to another
state $|S',m'>$
\begin{equation}
\frac{1}{\tau_{S,m}}=\sum_{(S',m') \ne (S,m)} p(S,m \rightarrow S',m')  
\end{equation}
which depends only on the additional parts of the Hamiltonian ${\cal
  H}_n$ which induce these transitions. In what follows we will try to
account for this lifetime assuming different possible interactions.

\subsection{Spin-Phonon Interaction}
To account for the temperature dependence of $T_1$ we should take into
consideration the transitions induced by spin-phonon interactions. The
spin-phonon coupling Hamiltonian \cite{Garg98PRL,Koloskova66}, ${\cal
H}_n$, can induce transitions between different spin states of the
molecule.
%
The transition rate from a state $|S,m>$ to a state $|S',m'>$ can be
calculated using the golden rule in perturbation theory
\cite{Hartmann96IJMP}
\begin{equation}
p(S,m \rightarrow S',m') = \frac{3}{2 \pi} \frac{|\left<
S,m|{\cal H}_{sp}|S',m' \right >|^2}{\hbar^4 \rho c^5} (E_{{\bf S},m}-E_{{\bf S}',m'})^3
\frac{1}{\exp \left[ (E_{{\bf S},m}-E_{ S',m'})/T \right]-1}
\end{equation}
This result involves the matrix element $\left< S,m|{\cal H}_{sp}|S', m'
\right>$ of the spin-phonon interaction, the phonon velocity $c$, the
specific mass $\rho$ and the energy difference $(E_{{\bf S},m}-E_{{\bf
S}',m'})$, where it was assumed that $E_{{\bf S},m} >E_{{\bf
S}',m'}$. To get the right order of magnitude, and simplify the
calculations, we assume a constant spin-phonon interaction matrix
element, arriving at
\begin{equation} \label{C}
p(S,m \rightarrow S',m') = \frac{C(E_{{\bf S},m}-E_{{\bf S}',m'})^3}{\exp \left[
(E_{{\bf S},m}-E_{{\bf S}',m'})/T \right]-1}
\end{equation}
where 
\[ C=\frac{3}{2 \pi} \frac{|\left< S,m|{\cal H}_{sp}|S',m' \right >|^2}{\hbar^4 \rho c^5}. \]

\begin{figure}[h]
\centerline{\epsfysize=6.0cm \epsfbox{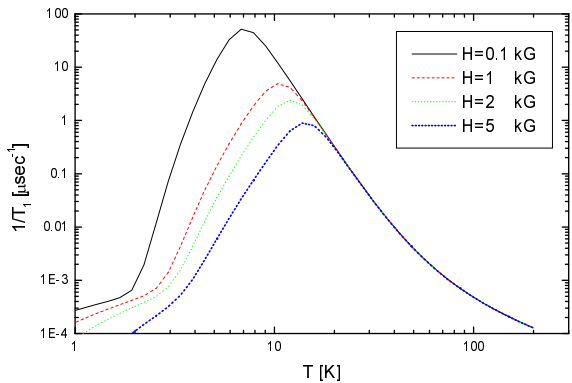}}
\caption{Spin lattice relaxation rate as a function of temperature for different magnetic fields in CrNi$_6$, when assuming spin-phonon interactions only. For this figure we used $C=400$ 1/sec K$^3$ and $A=5.2$ MHz in Eq. (\ref{T1calc}) and (\ref{C}).}\label{phonons}
\end{figure}

The transition probability due to spin-phonon interaction strongly
depends on temperature. At very low temperature phonons die out
exponentially with decreasing temperature, yielding a very low
transition probability, and extremely low spin lattice relaxation rate
values as seen in Figure~\ref{phonons}. Therefore this
interaction cannot give a full explanation to the nonzero spin lattice
relaxation rate at low temperatures, which is observed in experiments
\cite{Salman00,Salman02PRB}, and additional terms in the Hamiltonian
should be considered.

\subsection{Other Interactions}
In order to obtain a finite spin lattice relaxation rate at very low
temperatures, the lifetime of the levels should be finite. This cannot
be accounted for by spin-phonon interaction as seen in the previous
section. A finite lifetime can be achieved if one simply assumes a
finite broadening of the levels due to an additional interaction
${\cal H}_{int}$, giving a short lifetime $\tau_{int}$ for the levels.
The assumption implied by the experimental results \cite{Salman02PRB} is
that $\tau_{int}$ is temperature and field independent at low fields.

In this case the total lifetime of the levels consists of two
contributions, $\tau_{sp}$ due to spin-phonon interaction ${\cal
  H}_{sp}$ and $\tau_{int}$ due to the additional interaction ${\cal
  H}_{int}$, therefore
\begin{equation}
\frac{1}{\tau_{S,m}}=\left( \frac{1}{\tau_{sp}}+\frac{1}{\tau_{int}} \right).
\end{equation}
At high temperatures the value of the spin-phonon contribution to the
lifetime, $\tau_{sp}$, is much shorter than $\tau_{int}$ and the value
of $T_1$ is dominated by spin-phonon induced transitions, while at
very low temperatures $\tau_{sp}$ is much longer than $\tau_{int}$,
and the value of $T_1$ is dominated by $\tau_{int}$. At intermediate
temperatures both contributions to the spin lattice relaxation are
important.

\begin{figure}
\begin{center}
\begin{tabular}{cc}
\centerline{\includegraphics*[height=7.5cm,angle=-90]{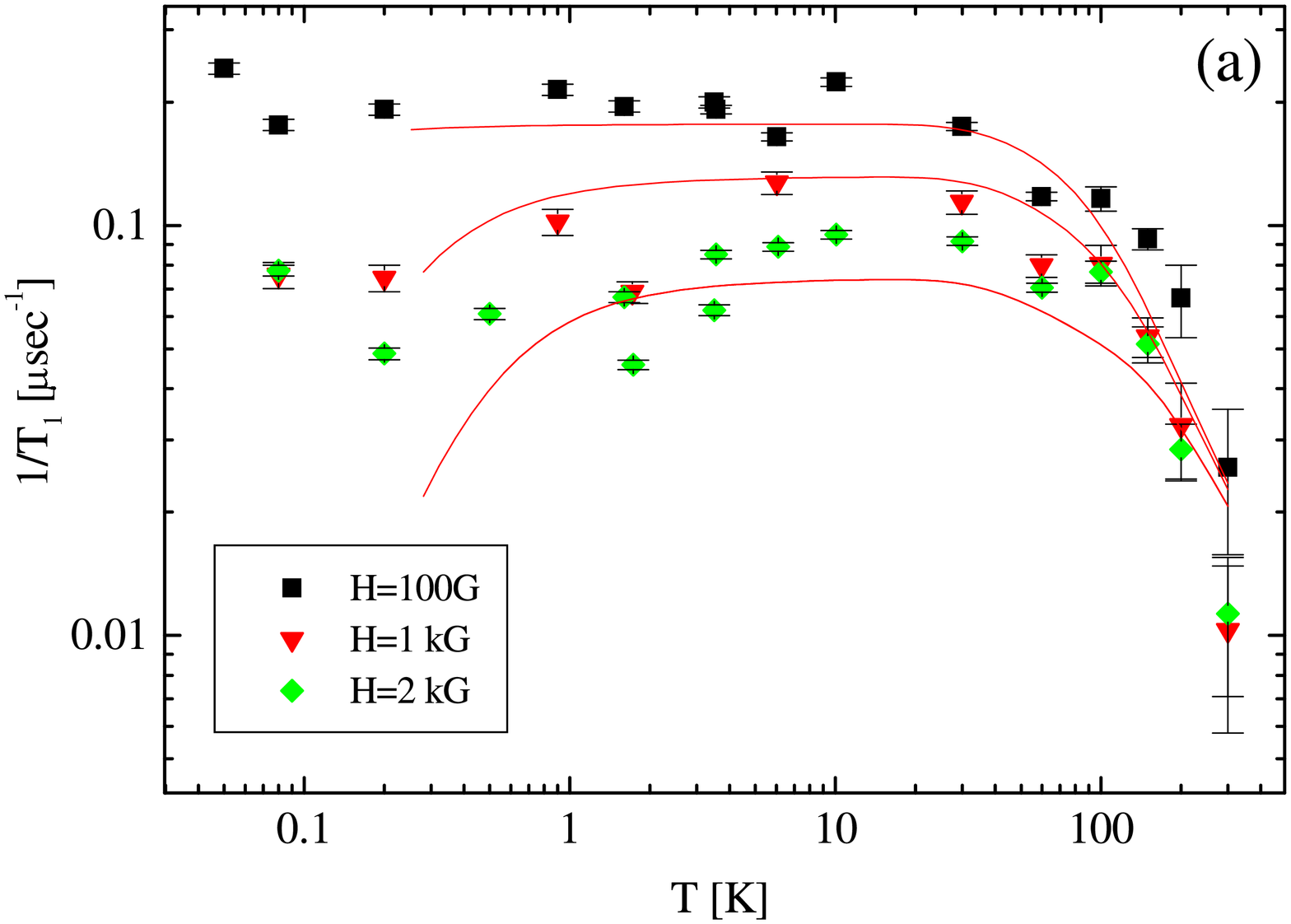}} \\
\centerline{\includegraphics*[height=7.5cm,angle=-90]{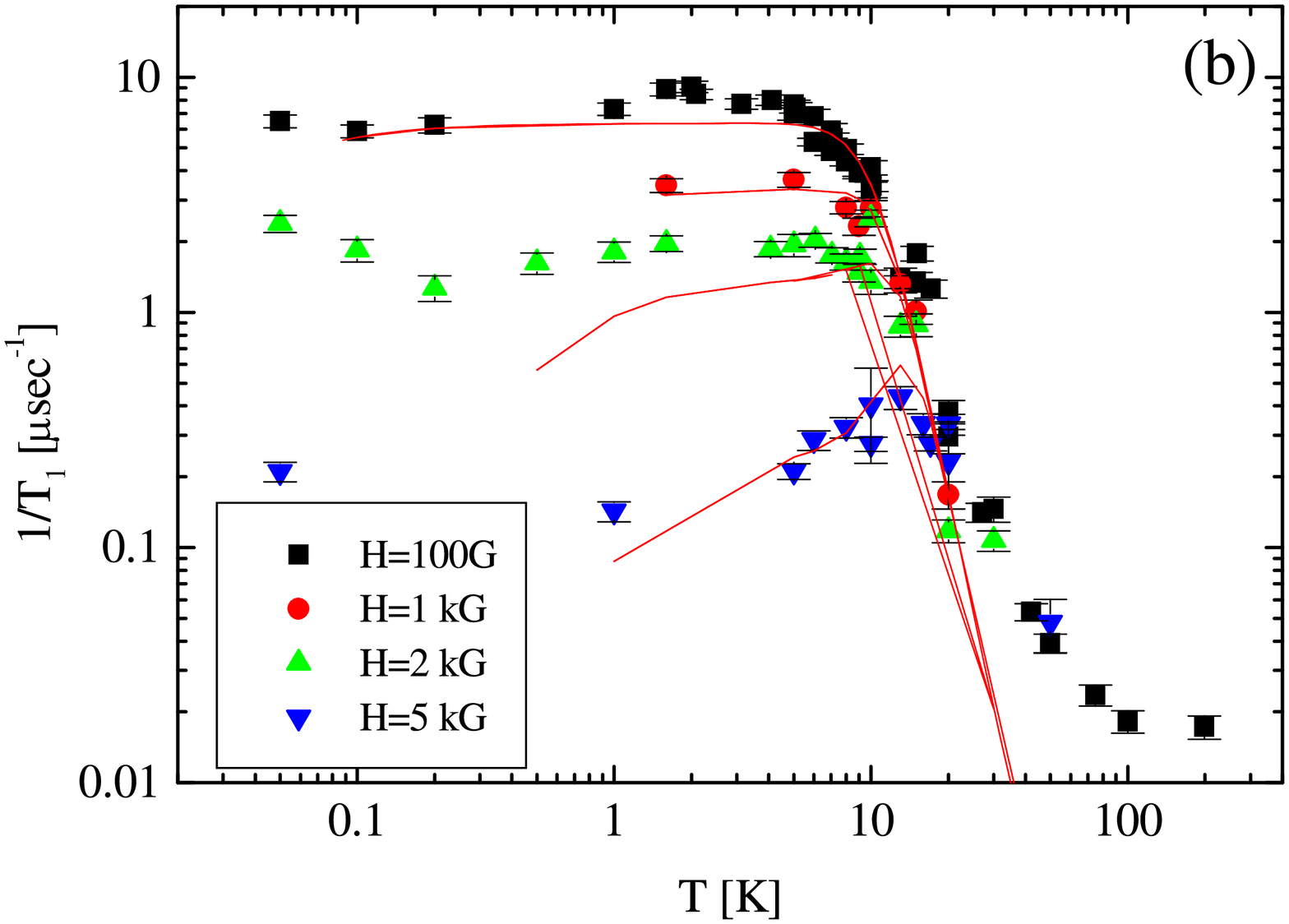}} \\
\centerline{\includegraphics*[height=7.5cm,angle=-90]{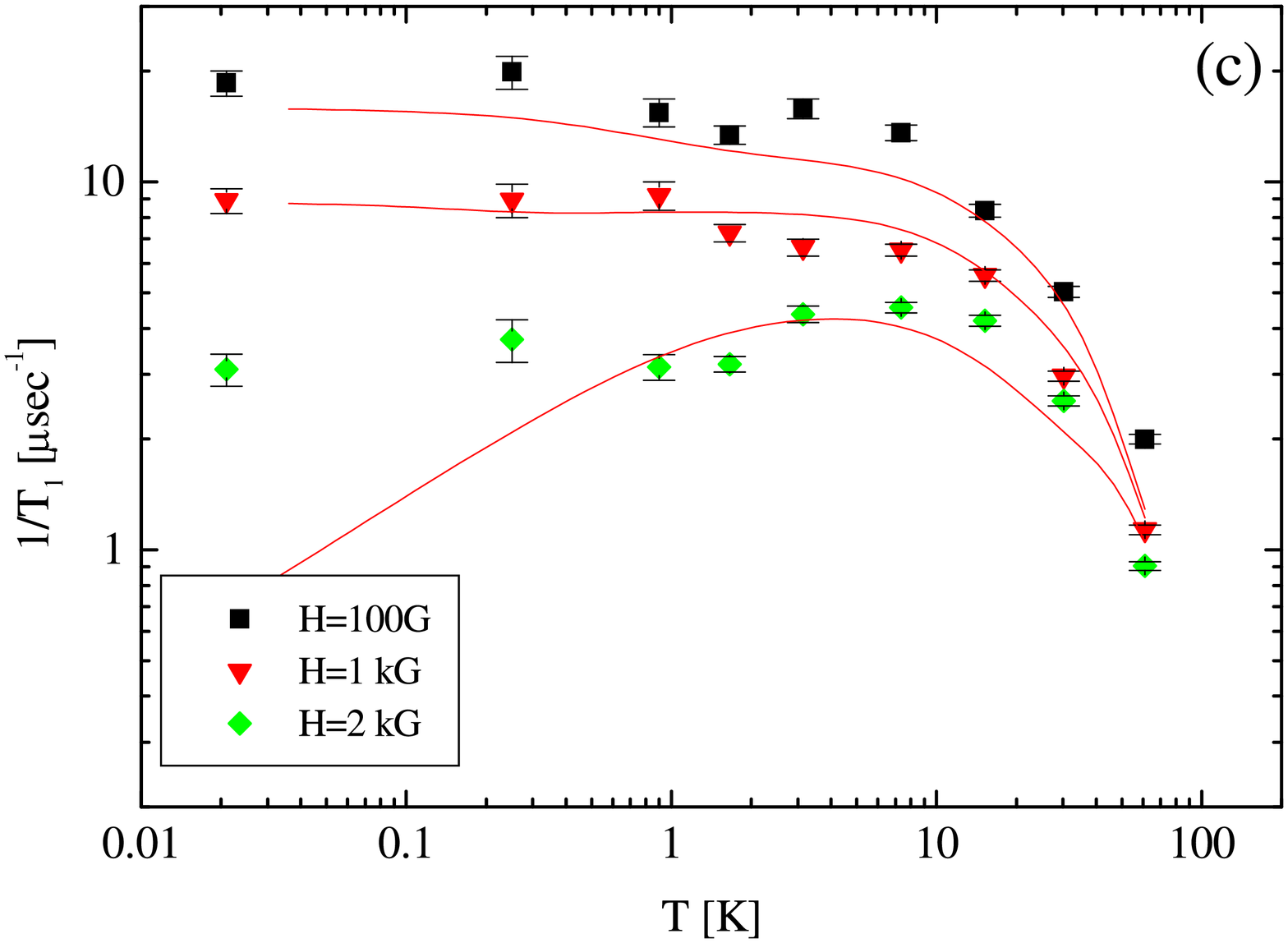}} 
\end{tabular}
\end{center}
\caption{The value of $1/T_1$ as a function of $T$ at different fields, measured using $\mu$SR in (a) CrCu$_6$, (b) CrNi$_6$ and (c) CrMn$_6$. The solid lines are fits to the calculated value (see text).} \label{LamvsT}
\end{figure}
In Figure~\ref{LamvsT} we present the experimental values of $1/T_1$,
measured by $\mu$SR \cite{Salman02PRB}, as a function of temperature
at different fields, for the molecules (a) CrCu$_6$, (b) CrNi$_6$, and
(c) CrMn$_6$. The solid lines are fits to the theoretical values
expected assuming transitions which are induced by spin-phonon
interaction in addition to ${\cal H}_{int}$. The fits give the parameters'
values summarized in Table~\ref{FitParams}.
\begin{table}[h]
\begin{center}
\begin{tabular}{|c||c|c|c|} \hline   
{\em Compound} &$\tau_{int}$ [nsec] & $C$ [1/sec K$^3$] & $A$ [MHz] \\ \hline \hline 
CrCu$_6$ & $7.0(8)$ & $0.13(9)$ & $1.7(1)$\\ \hline  
CrNi$_6$ & $11.0(8)$ & $400(60)$ & $5.2(2)$ \\ \hline  
CrMn$_6$ & $9.1(8)$ & $0.004(1)$ &$4.7(2)$ \\ \hline  \hline  
\end{tabular}
\end{center}
\caption{The fit parameters of the theoretical calculation of the spin lattice relaxation to the experimental values from $\mu$SR measurements} \label{FitParams}
\end{table}

The parameters quoted in Table~\ref{FitParams} were calculated using
Eq.~(\ref{T1calc}) with $\omega=\gamma_{\mu} H$ instead of $\omega'$,
since the experimental measurements were performed at low fields.  The
quoted values of $\tau_{int}$ indicate that ${\cal H}_{int}$ is of
order of $0.7-1.1$~K, which is larger than the Zeeman splitting in
fields up to $2$~kG, and self consistent with the use of
Eq.~(\ref{T1calc}) with $\omega=\gamma_{\mu} H$ instead of $\omega'$.

The fits in Figure~\ref{LamvsT} capture the essence of the temperature
and field dependence of $1/T_1$, considering the simplifications that
we have used, and it gives the correct general behavior of the
experimental data. However, at very low temperatures and high fields
the theoretical calculation deviates from the experimental
data. Similarly at high temperatures $T\gg J$, where $1/T_1$ is very
small, the theoretical calculation deviates from the experimental data
(especially in the case of CrNi$_6$). We believe that the origins of
the deviation at high temperatures is that the Hamiltonian
(\ref{Hfull}) does not describe the system well enough, and that the
value of $1/T_1$ is very small and is harder to estimate
experimentally.  This can also be seen in Figure~\ref{chit}, where at
high temperatures the calculated value of $\chi T$ deviates from the
experimental values.

\section{Conclusion}
The calculated spin lattice relaxation rate of a local probe, with
gyromagnetic ratio $\gamma$, in the isotropic HSM follows
Eq.~(\ref{T1calc}). This result remains valid assuming an isotropic
interaction between the probe's spin and the molecular spin, and
assuming a finite lifetime $\tau_{S,m}$ for the spin state $|S,m>$ at
all temperatures and fields. Eq.~(\ref{T1calc}) indicates that the
spin lattice relaxation rate in these molecules is independent of the
probe's gyromagnetic ratio at high magnetic fields, and therefore the
measured values of spin lattice relaxation by $\mu$SR and proton-NMR
scale at the same {\em external field} (at high fields), and not at
equal Larmor frequencies.


The calculated lifetime of the levels $\tau_{int}$ is found to vary
between $7-11$~nsec, which can be translated to a broadening of
$3-5$ mK. Therefore we expect that the interaction which introduces
the spin dynamics in these molecules produces level broadening of the
same order of magnitude.

The temperature and field independent levels broadening $1/\tau_{int}$
 calculated above can be attributed to an
interaction that does not commute with $S_z$ and induces transitions
between the different $m$ states. This interaction can be dipolar
between neighboring molecules, hyperfine between molecular and nuclear
spins, crystal field higher order terms, etc. However, the
striking fact is that the lifetime is similar in all three molecules,
indicating that it does not depend strongly on the spin of the
molecule or the coupling $J$ between ions inside the molecule, which
varies greatly between the three molecules.

This indicates that the weak dependence of the broadening on the spin
value cannot be explained by interactions which are quadratic in $S$
or have higher $S$ dependence. This rules out dipolar interactions
between neighboring molecules since in the three compounds the nearest
neighbor distance is $\sim 15$ \AA. Similarly, crystal field terms
which are allowed by the octahedral symmetry ($S^2$ or higher
\cite{Abragam86}) are unlikely.

The only mechanism suggested to date for level broadening of HSM,
which depends weakly on $S$ is the hyperfine interaction between
nuclear and electronic spins. This mechanism can account for the
finite spin lattice relaxation rate at very low temperatures
\cite{Waugh88PRB}. However, the values of broadening calculated above
might be inaccurate due to the simplifications made in the
calculations, but give the right order of magnitude expected from
hyperfine interactions \cite{Griffith64}.

Hyperfine interactions in anisotropic high spin molecules were studied
recently \cite{Caneschi98JMMM,Wernsdorfer99S}, and their effect on QTM
is becoming clearer \cite{Giraud01PRL,Giraud01condmat}. We believe
that this interaction also governs the spin dynamics of the isotropic
molecules at very low temperatures ($T<3$ K), while at high
temperature ($T>10$ K) the molecular spin dynamics are governed by
spin-phonon interactions.

\bibliographystyle{unsrt}
\bibliography{referances}

\newcommand{\noopsort}[1]{} \newcommand{\printfirst}[2]{#1}
  \newcommand{\singleletter}[1]{#1} \newcommand{\switchargs}[2]{#2#1}
\begin{thebibliography}{10}

\bibitem{Salman02PRB}
Z.~Salman, A.~Keren P.~Mendels V. Marvaud A. Scuiller M. Verdaguer J. S.~Lord 
  and C.~Baines.
\newblock {\em Phys. Rev. B}, 65:132403, 2002.

\bibitem{Lis80ACS}
T.~Lis.
\newblock {\em Acta Crystallogr. Sec. B}, 36:2042, 1980.

\bibitem{Thomas96N}
L.~Thomas et~al.
\newblock {\em Letters to Nature}, 383:145, 1996.

\bibitem{Freidman96PRL}
J.~Freidman et~al.
\newblock {\em Phys. Rev. Lett.}, 76:3830, 1996.

\bibitem{Sangregorio97PRL}
C.~Sangregorio et~al.
\newblock {\em Phys. Rev. Lett.}, 78:4645, 1997.

\bibitem{Ueda02PRB}
M.~Ueda, S.~Maegawa S.~Kitagawa.
\newblock {\em Phys. Rev. B}, 2002.

\bibitem{Lascialfari98PRL}
A.~Lascialfari, Z.~Jang F.~Borsa P.~Carretta and D.~Gatteschi.
\newblock {\em Phys. Rev. Lett.}, 81:3773, 1998.

\bibitem{Politi95PRL}
P.~Politi, A.~Retori F.~Hartmann-Boutron and J.~Vilain.
\newblock {\em Phys. Rev. Lett.}, 75:537, 1995.

\bibitem{Barra97PRB}
A.~L.~Barra D.~Gatteschi and R.~Sessoli.
\newblock {\em Phys. Rev. B}, 56:8192, 1997.

\bibitem{Barra98JMMM}
A.~Barra, A.~Caneschi D.~Gatteschi and R.~Sessoli.
\newblock {\em JMMM}, 177, 1998.

\bibitem{Fort98PRL}
A.~Fort, A.~Rettori J.~Villain-D.~Gatteschi and R.~Sessoli.
\newblock {\em Phys. Rev. Lett.}, 80:612, 1998.

\bibitem{Caneschi98JMMM}
A.~Caneschi, T.~Ohm D.~Rovai-C.~Sangregorio and R.~Sessoli.
\newblock {\em JMMM}, 177, 1998.

\bibitem{Villain94EL}
J.~Villain et~al.
\newblock {\em Europhys. Lett.}, 27:159, 1994.

\bibitem{Prokofev98PRL}
N.~Proko\'fev and P.~Stamp.
\newblock {\em Phys. Rev. Lett.}, 80:5794, 1998.

\bibitem{Salman00}
Z.~Salman, A.~Keren P.~Mendels-A.~Scuiller and M.~Verdaguer.
\newblock {\em Physica B}, 106:289, 2000.

\bibitem{Garanin97PRB}
D.~A. Garanin and E.~M. Chudnovsky.
\newblock {\em Phys. Rev. B}, 56:11102, 1997.

\bibitem{Garg98PRL}
A.~Garg.
\newblock {\em Phys. Rev. Lett.}, 81:1513, 1998.

\bibitem{Miyashita01condmat}
S.~Miyashita and N.~Nagaosa.
\newblock {\em cond-mat/0108063}, 2001.

\bibitem{White}
R.~White.
\newblock {\em Quantum Theory of Magnetism}.
\newblock Springer-Verlag, second edition, 1983.

\bibitem{Koloskova66}
N.~Koloskova.
\newblock In A.~A. Manenkov and R.~Orbach, editors, {\em Spin Lattice
  Relaxation in Ionic Solids}, page 232. Harper and Row, 1966.

\bibitem{Hartmann96IJMP}
P.~Politi F.~Hartmann-Bourtron and J.~Villain.
\newblock {\em Int. J. Mod. Phys.}, 10:2577, 1996.

\bibitem{Abragam86}
B.~Bleaney A.~Abragam.
\newblock {\em Electron Paramagnetic Resonance of Transition Ions}.
\newblock Dover Publications: Mineola, 1986.

\bibitem{Waugh88PRB}
J.~S. Waugh and C.~P. Slichter.
\newblock {\em Phys. Rev. B}, 37:4337, 1988.

\bibitem{Griffith64}
J.~S. Griffith.
\newblock {\em The Theory of Transition-Metal Ions}.
\newblock Cambridge University Press, 1994.

\bibitem{Wernsdorfer99S}
W.~Wernsdorfer and R.~Sessoli.
\newblock {\em Science}, 284:133, 1999.

\bibitem{Giraud01PRL}
R.~Giraud, W.~Wernsdorfer A.~M. Tkachuk D.~Mailly and B.~Barbara.
\newblock {\em Phys. Rev. B}, 87:57203, 2001.

\bibitem{Giraud01condmat}
R.~Giraud, W.~Wernsdorfer A.~M. Tkachuk D.~Mailly and B.~Barbara.
\newblock {\em cond-mat/0108133}, 2001.

\end{thebibliography}
\end{document}